\documentclass[12pt,preprint]{aastex}

\shorttitle{emerging Active Region, interaction, and fan-shaped surge}
\shortauthors{Zhong et al.}

\begin{document}

\title{The dynamics of AR 12700 in its early emerging phase I:  interchange reconnection }

\author{Sihui Zhong\altaffilmark{1,2}, Yijun Hou\altaffilmark{1,2},
        and Jun Zhang\altaffilmark{1,2}}

\altaffiltext{1}{CAS Key Laboratory of Solar Activity, National Astronomical Observatories,
Chinese Academy of Sciences, Beijing 100101, China; zsh@nao.cas.cn; zjun@nao.cas.cn}

\altaffiltext{2}{University of Chinese Academy of Sciences, Beijing 100049, China}

\begin{abstract}
The emergence of active regions (ARs) leads to various dynamic activities.
Using high-resolution and long-lasting H$\alpha$ observations from the New Vacuum Solar Telescope, we report the dynamics of NOAA AR 12700 in its emerging phase on 26 February 2018 in detail.
In this AR, constant interchange reconnections between emerging fibrils and preexisting ones were detected.
Driven by the flux emergence, small-scale fibrils observed in H$\alpha$ wavelength continuously emerged at the center of the AR and reconnected with the ambient preexisting fibrils, forming new longer fibrils. We investigate three scenarios of such interchange reconnection in two hours. Specially, the third scenario of reconnection resulted in the formation of longer fibrils that show pronounced rotation motion.
To derive the evolution of the magnetic structure during the reconnections, we perform nonlinear force-free field extrapolations. The extrapolated three-dimensional magnetic fields clearly depict a set of almost potential emerging loops, two preexisting flux ropes at 03:00 UT before the second reconnection scenario, and a set of newly formed loops with less twist at 03:48 UT after the third reconnection scenario. All of these extrapolated structures are consistent with the fibrils detected in H$\alpha$ wavelength.
The aforementioned observations and extrapolation results suggest that the constant interchange reconnections resulted in that the magnetic twist was redistributed from preexisting flux ropes towards the newly-formed system with longer magnetic structure and weaker twist.
\end{abstract}


\section{Introduction}
The emergence of active regions (ARs) is of fundamental importance in solar physics. Observations of how ARs emerge reveal
the transport processes that bring magnetic fields to the solar atmosphere. The emergence of ARs is a multi-stage process \citep{2015LRSP...12....1V}.
Initially, toroidal magnetic fields are generated close to the base of the convection zone. Then, presumably triggered by deep convective flows
and buoyant instabilities, magnetic flux tubes rise towards the surface as $\Omega$-shaped loops, break through it, and leave footprints
in the forms of sunspots and plages \citep{1987ARA&A..25...83Z, 1997smf..conf....3M, 2007ASPC..369..335M, 2009LRSP....6....4F}.
The evolution of emerging flux tubes from below the solar surface to the corona is associated with various phenomena such as
moving magnetic features (MMF; \citealt{2002ApJ...566L.117Z,2003A&A...399..755Z}), plages, Ellerman bombs (EBs; \citealt{2013MmSAI..84..436N}),
arch filament system (AFS), micro-pores \citep{2017A&A...600A..38G}, rotational bipoles \citep{2009ApJ...697.1529F,2013SoPh..282..503K},
and jets/surges \citep{2014ApJ...794..140V}. The relationships between the aforementioned phenomena and flux emergence are shown in the review of \cite{2014SSRv..186..227S}.
Recent high-resolution observations of small-scale emergence events give insight to how ARs appear on the solar surface.
\cite{2011PASJ...63.1047O} studied the nature of flux emergence with Hinode \citep{2007SoPh..243....3K}/Solar Optical Telescope (SOT; \citealt{2008SoPh..249..167T}) data. \cite{2012ApJ...759...72C} presented the naked emergence of ARs observed by the \emph{Solar Dynamics Observatory }(\emph{SDO}; \citealt{2012SoPh..275....3P}).

The emergence of ARs leads to various dynamic activities.
Observations and numerical simulations have shown that the interaction of newly emerging magnetic flux with pre-existing magnetic fields leads to coronal heating \citep{1991saaj.conf..169S,2002mwoc.conf...39M,2004ESASP.575..241P,2005A&A...439..335G} and redistribution of helicity \citep{2001ApJ...561..406Z,2003ApJ...584..479Z}.
When magnetic flux emerges from beneath the photosphere, it may reconnect with the preexisting fields. Interchange is one model of reconnection, which often occurs between closed and open fluxes \citep{2002JGRA..107.1028C}.
Here we define the interchange reconnection (IR) as a process that two sets of magnetic loops interact with each other and interchange their footpoints.
Observational evidences supportive of IR between emerging active region and coronal hole (CH) include corona dimming
\citep{2007AN....328..773B} and the retreat of the CH boundary \citep{2018ApJ...863L..22K}.
\cite{2014A&A...570A..93L}
reported the detailed interchange reconnection process as a way to convert mutual helicity to self-helicity
by employing observations from the \emph{Interface Region Imaging Spectrometer} (\emph{IRIS}; \citealt{2014SoPh..289.2733D}).

Numerical simulations of AR emergence bring insight into the magnetic and dynamic properties of the emergence process.
Recent three-dimensional (3D) megnetohydrodynamics (MHD) simulations are able to produce an AR based on different emergence conditions (\citealt{2012A&A...537A..62A, 2014ApJ...785...90R, 2015ApJ...811..138T, 2017ApJ...846..149C} and references therein).
It is noteworthy that, the model of \cite{2010ApJ...720..233C} has rather successfully explained some observational properties associated with ARs emergence, including elongated granules, mixed polarity patterns in the emergence zone, pore formation, and light bridges.
In addition, flux-emergence experiments often see the magnetic reconnection between emerging magnetic flux and the preexisting ambient fields.
\cite{2010ApJ...714..517E} investigated the effect of IR on the dynamics and topology of coronal hole boundaries. Based on 3D MHD calculations, \cite{2012SSRv..172..209E} argued that IR plays a defining role in the evolution of the coronal magnetic field, and therefore the generation of the slow solar wind.

Despite preexisting MHD models for explaining AR emergence, observational evidences of the detailed emergence process have rarely been reported.
Previous studies focused more on photospheric layer and coronal response of the emergence events.
\cite{2009ApJ...697..913O} reported the emergence of a flux rope at the polarity inversion line (PIL) in AR 10953, which was controversial.
\cite{2010ApJ...716L.219M} constructed a dynamic flux emergence model and found that the signatures of \cite{2009ApJ...697..913O} are not sufficient to uniquely identify an emerging flux rope. \cite{2012SoPh..278...33V} argued that the emergence of the flux rope did not take place at the PIL.
\cite{2017ApJ...845...18Y} observed a small-scale emerging flux rope near a large sunspot and the entire process from its emergence to eruption by Big Bear Solar Observatory (BBSO)/Goode Solar Telescope (GST; \citealt{2012ASPC..463..357G}).
In addition, limited by the low resolution of previous observations, distinct detections of interchange reconnection in the emerging ARs are rare.
In the present work, using the high-resolution and long-lasting H$\alpha$ observations acquired at the New Vacuum Solar Telescope (NVST; \citealt{2014RAA....14..705L}) and the simultaneous observations from the \emph{SDO}, we present the detailed processes of three scenarios of interchange reconnection in AR 12700 on 26 February 2018.
It is noteworthy that the H$\alpha$ observations we adopted last for 5 hours, covering the early emerging phase of AR 12700.
These observations provide a complete view of IR, as they cover all the atmospheric layers from the photosphere to the corona at high temporal and spatial resolution.
In particular, the H$\alpha$ observations clearly depict the emergence of flux tubes, and rotational motion of fibrils which are formed via reconnection between the emerging flux tubes and the preexisting fibrils.
Moreover, the results derived from the nonlinear force-free field (NLFFF) extrapolations are consistent with the observations, providing more details on the changes of magnetic structures during the IRs.

Our paper is organized as follows. Section 2 describes the observations and data analysis taken in our
study. In Section 3, we investigate three scenarios of IR between
emerging flux and preexisting field in great detail.
Finally, we summarize the major findings and discuss the results in Section 4.

\section{Observations and Data Analysis}
On 26 February 2018, NOAA AR 12700 emerged with $\beta-$configuration at solar disk location N04W01.
The NVST was pointed at this region on 26 February, and
one series of H$\alpha$ 6562.8 {\AA} observations were taken from 02:01:00 UT to 06:56:00 UT with
a cadence of 8 s, a field of view (FOV) of 152{\arcsec} $\times$ 151{\arcsec}, and a spatial sampling
of 0.{\arcsec}136 pixel$^{-1}$.
These H$\alpha$ observations clearly reveal the detailed emergence process in chromosphere, including fibrils emergence, interactions between different groups of fibrils, and untwisting motion of fibrils. 

Moreover, we have also analyzed the data taken by the Helioseismic and Magnetic Imager (HMI; \citealt{2012SoPh..275..207S}) and Atmospheric Imaging Assembly (AIA; \citealt{2012SoPh..275...17L}) on board the \emph{SDO} to figure out the photospheric magnetic field evolution and coronal response during the emergence of AR 12700.
The HMI data adopted here were obtained from 2018 February 25 to 26, with a cadence of 45 s and a pixel size of ~0.{\arcsec}5.
The AIA provides successive full-disk images of the multi-layered solar atmosphere with ten passbands, seven of which are in extreme ultraviolet (EUV) channel and observed with a cadence of 12 s and a pixel size of 0.{\arcsec}6.
Here we focus on the 171 {\AA} wavelength, which manifests the coronal brightenings clearly.
All the data are calibrated with standard solar software routines, and all images observed by the \emph{SDO} are differentially rotated to a reference time (04:00:00 UT on 26 February). Moreover, data from all telescopes and instruments are carefully co-aligned, and the region of interest (ROI) is spatially and temporally extracted from the different channels.

In order to investigate the evolution of the magnetic structures during the reconnections, we perform NLFFF extrapolations
at 03:00 UT and 03:48 UT on 26 February with HMI photospheric vector magnetic fields as the boundary condition.
The extrapolations use the ``weighted optimization" method \citep{2004SoPh..219...87W, 2012SoPh..281...37W}.
The vector magnetograms are preprocessed by a procedure developed by \cite{2006SoPh..233..215W} to remove the force and noise.
Both NLFFF extrapolations are performed in a box of 288 $\times$ 168 $\times$ 256 uniformly spaced grid points (104 $\times$ 61 $\times$ 93 Mm$^{3}$).

\section{Results}
AR 12700 emerged near the center of the solar disk on 26 February 2018 , which is shown in Figure 1.
Prior to the
NVST H$\alpha$ observations, sequence of HMI magnetograms show remarkable rotation of magnetic patches and separation
of the fields with opposite polarities in this AR. At 22:04:10 UT on 25 February, the negative patch denoted by white
contour in Figure 1(a) owned an elongated shape and a 19$^{\circ}$ angle between its main axis and the horizontal direction.
Then, the elongated patch rotated anticlockwise, increasing the angle up to 166$^{\circ}$ at 00:55:55 UT on 26 February.
The mean rotating speed was about 0.85$^{\circ}$ min$^{-1}$. In addition, a bipole (marked by the white brackets
in panel (b)) emerged at 23:05:40 UT on 25 February and its positive patch (denoted by the red triangle) shifted
northeastward with a velocity of 0.5 km s$^{-1}$ in the following one hour.

The H$\alpha$ observations reveal that constant interchange reconnections occurred in the central region of AR 12700, which is
approximately outlined by the blue dashed rectangle in Figure 1(a) and extended in Figures 2-5.
From 02:00 UT
to 04:00 UT on 26 February, three scenarios of IR were detected occurring between the emerging fibrils
and the preexisting ones. Figure 2 (also see animation 1) shows the first scenario of reconnection which occurred
from 02:12 UT to 02:32 UT. The H$\alpha$ observations (panels (a1)-(a3)) clearly show that two groups of chromospheric
fibrils, i.e., preexisting fibrils (PF) and emerging fibrils (EF), successively interacted with each other from 02:24 UT
to 02:30 UT. As a result, a new group of fibrils (NF) were formed (see panel (a3)).
The corresponding
HMI magnetograms reveal that EF and PF were rooted in magnetic fields with opposite polarities, the southwest
footpoint of EF was rooted in the main negative polarity of the AR and its northeast footpoint was located in
the positive fields emerging between the main polarities.
After interaction, the west leg of NF was close to the EF's southwest footpoint and its east leg was close to
PF's northeast footpoint (see panels (b) and (d)). It is consistent with the condition of IR. Moreover, AIA 171 {\AA} observations revealed brightenings
appearing at the intersection of PF and EF, and lasting from 02:21 UT to 02:32 UT, which further implies that
the IR occurred between EF and PF.

In Figure 3 (also see animation 2), the second reconnection scenario is displayed, which occurred from 03:00 UT to 03:17 UT.
At 03:02:09 UT, there were three groups of chromospheric fibrils: small-scale emerging fibrils (EF: EF1 and EF2) and
two groups of preexisting fibrils (PF1 and PF2). The magnetic connections of EF and PF1 (see panel (b1)) here
are similar to that shown in Figure 2(b). The western legs of EF and PF1 were rooted in the main negative polarity of AR 12700.
Interactions between EF and PF1 started at 03:02 UT, continued for about 15 minutes, and then a new group of fibrils (NF1) were formed.
As shown in panel (b2), NF1 connected two main polarities of the AR, with its west leg close to EF's southwest footpoint
and its east leg close to PF1's northeast footpoint. The changes of the magnetic connections of these two groups of fibrils are representative IR signatures.

Figure 4 displays the emergence of the small-scale fibrils and its associated thermal properties at the onset of the second reconnection.
The newly emerging fibrils were clearly observed at 03:03 UT (see panel (b2)).
During the EF emergence, brightening in 171 {\AA} channel appeared in EF, the west footpoint of PF1, and the intersections of PF1 and EF, lasting from 03:03 UT to 03:06 UT.
Note that the brightenings were first observed in north part of EF and then in the south part (see panels (a1)-(a4)).
To investigate the detailed process of the EF emergence, we make time slices (panels (c1)-(c2)) in H$\alpha$ channel along vertical cut ``A-B'' and slit ``C-D'' shown in panel (b1).
Panel (c1) shows that EF initially rose at a projected velocity of 13 km s$^{-1}$, which is comparable to the previous studies \citep{1988ApJ...333..420C, 2014LRSP...11....3C}. EF's rising projected height was 1.5 Mm. At 03:03 UT, H$\alpha$ brightenings began appearing at the two sides of EF.
The light curve of H$\alpha$ superposed in panel (c2) shows that the average emission strength peaked around 03:05:30 UT.
Significant brightenings in H$\alpha$ channel and EUV channels at the interaction sites between EF and PF1
indicate the occurence of reconnection.

The third reconnection is shown in Figure 5 (also see animation 2).
From 03:18 UT to 03:34 UT, PF2 was split into two groups, one group interacted with EF and led to the formation of new longer fibrils similar to NF1 which showed a pronounced rotation motion from 03:24 UT to 03:33 UT. Note that the rotation originated from the intersection between EF and PF2. At 03:34 UT, the southernmost part of EF was lifted and interacted with the other group of PF2, leading to the formation of another group of fibrils.
As a result, several groups of newly-formed longer fibrils (NFs) are produced to connect the main polarities after the constant reconnections.
Meanwhile, brightenings (denoted by the green contour and arrows) in 171 {\AA} channel appeared in the intersections of different groups of fibrils, peaked around 03:31 UT (panel (b2)), and lasted from 03:27 UT to 03:39 UT.
The corresponding HMI magnetograms (panel (c)) shows that the footpoints of NFs were anchored in the locations of the EF's negative footpoint and PF2's positive footpoint,
suggesting the magnetic connections are consistent with IR model.

Based on the photospheric vector magnetic fields at 03:00 UT and 03:48 UT, we extrapolate the 3D structure of the target AR by using NLFFF modelling.
For visualizations of the emerging fibrils and preexisting ones in the second and third reconnection scenarios mentioned above, we select a region with the
FOV similar to Figure 1 from the NLFFF extrapolations and display the results in Figure 6.
Figures 6(a) and 6(b)
show the extrapolation results at 03:00 UT from the top view and side view, respectively.
The emerging loops (EL)
were overlaid by two magnetic flux ropes (FR1 and FR2). The north part of EL (EL(N)) is higher than its south part (EL(S)).
At 03:48 UT, a set of longer loops (NL) with weak twist were formed. The modelling results are consistent with the
observations shown in Figures 3--5, for that EL(N), EL(S), FR1, FR2, and NL here is corresponding well to EF1, EF2, PF1, PF2, and NFs
in H$\alpha$ images, respectively. In addition, the extrapolations show that the twist angle of FR1, FR2 is about 2$\pi$ and 4$\pi$, respectively.

\section{Summary and Discussion}
In this paper, we study the dynamics of AR 12700 in its emerging phase on 26 February by using observations from the NVST and \emph{SDO}.
The photospheric evolution of the emerging AR is characterized by the rotation of magnetic patches and separation of emerging bipoles.
Driven by the flux emergence,
IR between emerging flux and preexisting fields constantly occurred in the upper atmosphere in AR 12700.
At the centre of this AR, we investigate three such processes that small-scale emerging fibrils reconnect with the overlying preexisting ones, accompanied by H$\alpha$ and EUV brightening, finally forming new groups of fibrils.
Specially, during the third reconnection scenario, the formation of longer fibrils via reconnection shows remarkable rotation motion.
Besides, the extrapolated 3D fields
clearly depict the small-scale emerging loops, two overlying flux ropes, and the newly-formed loops. They coincide well with the observed emerging fibrils, two groups of preexisting fibrils, and the newly-formed longer fibrils, respectively.

The emergence of ARs is associated with various dynamic activities, such as jets and flares which are triggered by reconnections \citep{2007Sci...318.1591S, 2013A&A...559A...1S, 2014IAUS..300..184A}. During the emergence of AR 12700, we detect IRs and surge-like activities. Surge-like activities will be investigated in another upcoming paper. In the present work, three processes of IRs are investigated in detail. We confirm these reconnections are IR for the solid evidences as follows: (1) distinct interactions between constant emerging fibrils and preexisting fibrils,
(2) brightenings in H$\alpha$ and EUV wavelengths at their intersections, and (3) the newly-formed longer fibrils due to the interactions between two groups of H$\alpha$ fibrils that show changes of magnetic connections.

The NLFFF modelling reveals that EL is almost potential, with its north part (EL(N)) higher than its south part (EL(S)). This coincides well with the observations shown in Figures 4(a1)-(a4), that is, brightenings in 171 {\AA} channel first appeared in north part of EF and then in the south part of EF. Considering the extrapolation results and Figures 3-4,
we suggest that PF1 firstly reconnected with EF1, which was higher and closer to PF1 at 03:00 UT, and then reconnected with EF2, which rose up to the height of PF1 at 03:03:33 UT.
As shown in Figures 6(c)-(d) and animation 2, the third reconnection scenario finally resulted in the formation of the twisted structure NL with weaker twist than FR1 and FR2.
Similar observations have been reported by \cite{2016NatCo...711837X}. They investigated the rotational motion of the erupted filament enabled by the reconnection with the chromospheric fibrils and proposed that the reconnection between the filaments and less twisted flux leads to the release of twist.
Using BBSO/GST, \cite{2017A&A...603A..36K} found that reconnection of cool loops caused the formation of an unstable flux rope that showed counterclockwise rotation,
which was driven by the rapid flux cancellation in decaying phase of AR 12353.
However, in our case, the reconnections were driven by flux emergence in AR emerging phase.

According to the twisted threads of NL shown in Figures 6(c)-(d), we estimate its twist angle is less than 2$\pi$. As mentioned above, the twist angle of FR2 is about 4$\pi$.
These extrapolation results suggest that the IR between potential emerging flux and preexisting flux ropes resulted in that the
magnetic twist was redistributed from the preexisting flux ropes to the newly-formed system with longer
magnetic structure and weaker twist. As magnetic flux emergence going on, the reconnections reconfigure
the magnetic fields within the AR, and redistribute the magnetic twist.
Similar scenarios have been proposed by \cite{1996ApJ...473..533P} and \cite{1998SoPh..182..145C} with data of about 1{\arcsec} pixel size. In the present work, the observations with a 0.{\arcsec}136 pixel size enable us to identify the details throughout the reconnection process.
The result revealed in our work is
different from the previous emerging-reconnection picture that the emerging fields are twisted and then the
twist is transported to the newly formed structure via reconnection with the potential overlying preexisting
fields \citep{2003ApJ...593.1217P,2004ApJ...609.1123F}. According to the work of \cite{2016NatCo...711837X},
when filaments reconnect with less twisted flux, the twist tends to equilibrate along the new structure,
resulting from a true propagation of twist from the more twisted to the less twisted part (as a torsional Alfv{\'e}n wave packet). During the third reconnection scenario in our work, reconnections between FR2
with twist angle of 4$\pi$ and the potential flux result in NL whose twist angle is less than 2$\pi$.
Such process
of twist propagation presents as an rotational motion of the newly-formed longer fibrils.

\acknowledgments{
The authors would like to thank the anonymous referee for his or her thoughtful and helpful comments. The data are used courtesy of \emph{NVST} and \emph{SDO} science teams. \emph{SDO} is a mission of
NASA's Living With a Star Program.
This work is supported by the National Natural Science Foundations of China (11533008, 11790304, 11773039, 11673035, 11673034, 11873059, and 11790300), and Key Programs of the Chinese Academy of Sciences (QYZDJ-SSW-SLH050).
}

{}
\clearpage

\begin{figure}
\centering
\includegraphics [width=0.96\textwidth]{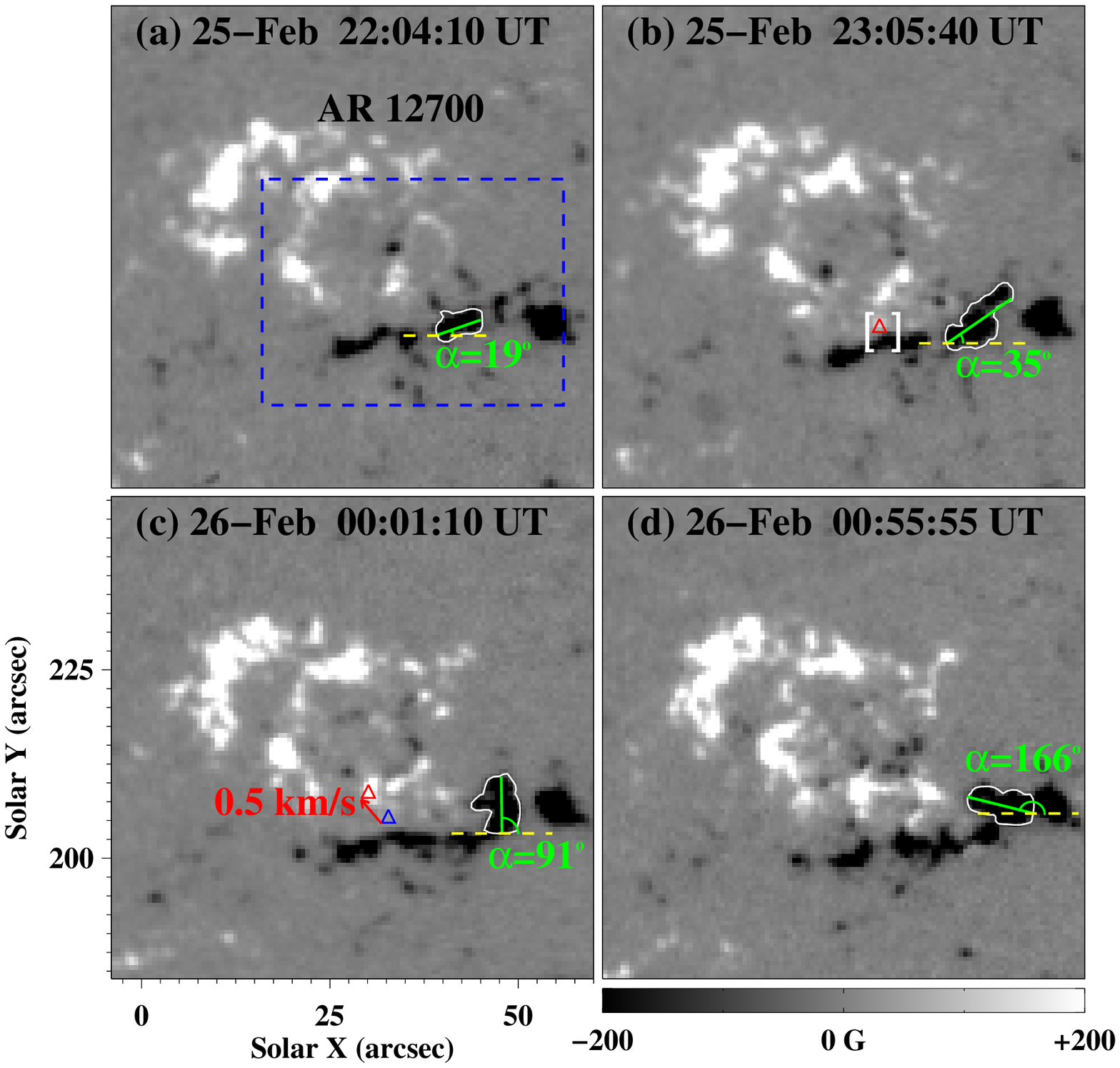}
\caption{Sequence of \emph{SDO}/HMI magnetograms displaying the evolution of AR 12700 from 22:00 UT on 25 February to 01:00 UT on 26 February 2018.
The white contours outline the magnetic patch with negative polarity, which shows obvious rotation motion. The green solid line shows the axis of this patch, the yellow dashed line represents the horizonal direction, and the angle between them is $\alpha$.
The blue dashed rectangle approximates the FOV of Figures 2-5.
The white square brackets in panel (b) highlight a newly-emerging bipole, and the red triangle marks the positive magnetic element of the bipole.
Triangles in panel (c) play the same role, with the red one indicating the  present location and the blue one indicating the previous location in panel (b).
}
\label{fig1}
\end{figure}

\begin{figure}
\centering
\includegraphics [width=0.96\textwidth]{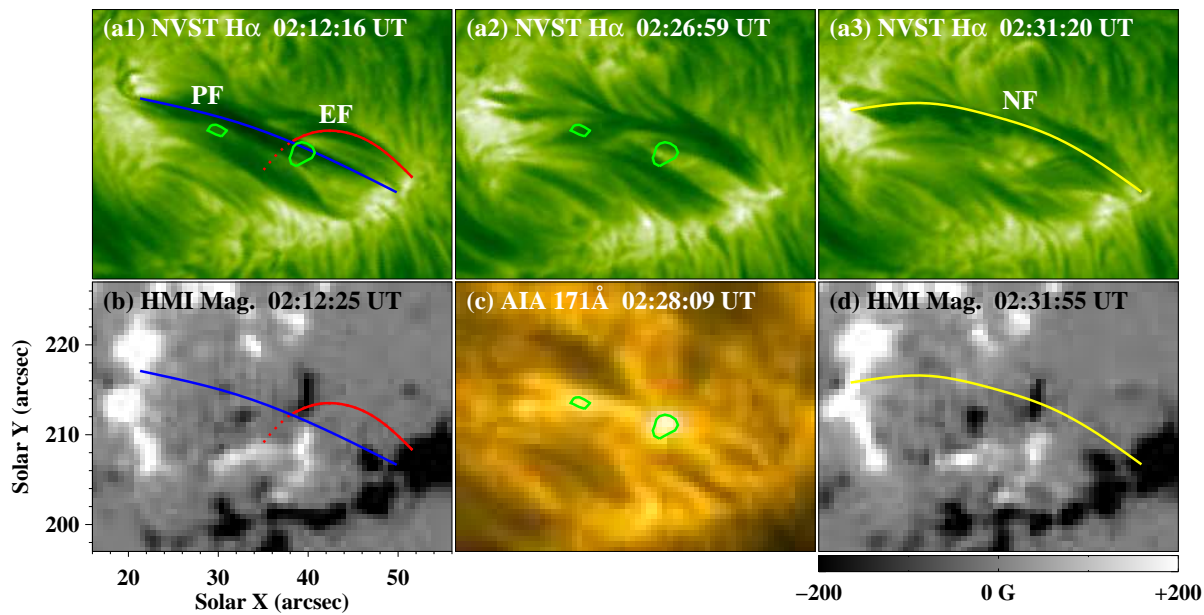}
\caption{The first scenario of interchange reconnection between emerging fibrils and preexisting ones observed from 02:12 UT to 02:31 UT on 26 February.
Panels (a1)-(a3): NVST H$\alpha$ images displaying the scenes before, during, and after the reconnection. The blue and red curves indicate the preexisting fibrils (PF) and the newly-emerging small ones (EF),
which are duplicated in panel (b). The yellow curve indicates the newly-formed fibrils (NF) resulting from the reconnection, which is overplotted on panel (d).
Panels (b)-(d): \emph{SDO}/HMI magnetograms and AIA 171 {\AA} image corresponding to the top panels. The green contours mark the locations of AIA 171 {\AA} brightenings, which are overplotted on panels (a1)-(a2).
Online animation (movie1.mov) displays NVST H$\alpha$ and \emph{SDO}/AIA 171 {\AA} images shown in Figure 2. The 10s animation covers ~29 minutes from 02:05 UT to 02:34 UT on 26 February 2018.
}
\label{fig2}
\end{figure}

\begin{figure}
\centering
\includegraphics [width=0.85\textwidth]{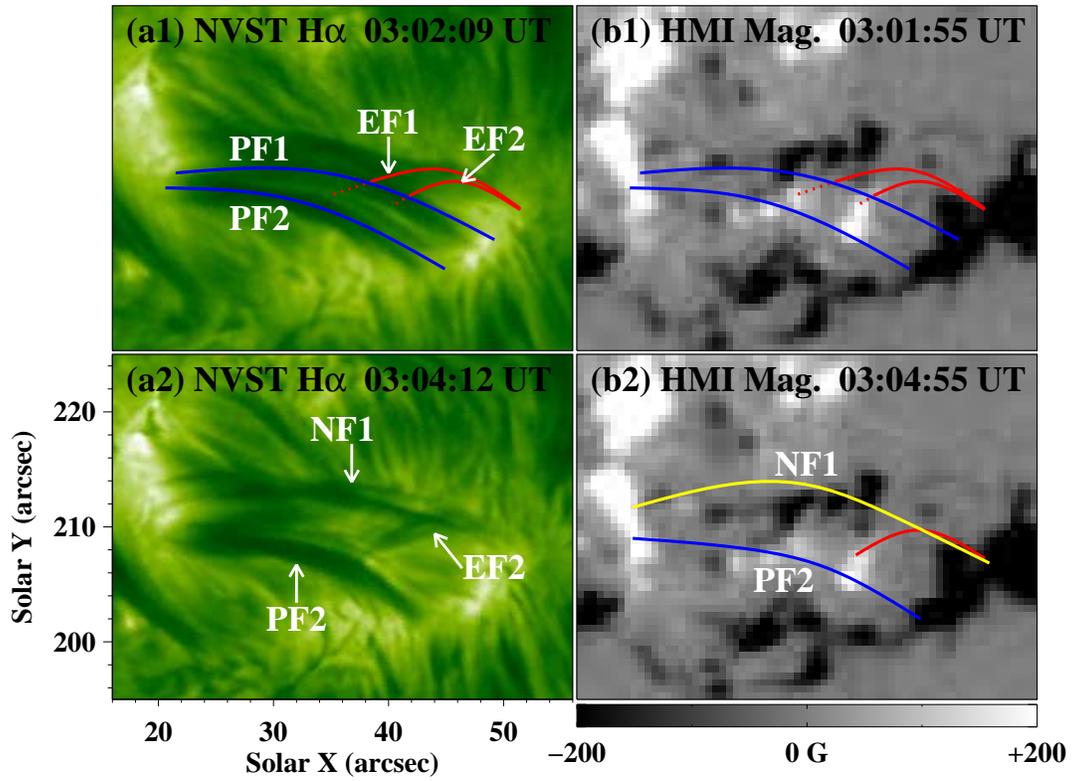}
\caption{NVST H$\alpha$ and \emph{SDO}/HMI observations displaying the second reconnection scenario from 03:00 UT to 03:17 UT. The features marked here are similar to those in Figure 2.
}
\label{fig3}
\end{figure}

\begin{figure}
\centering
\includegraphics [width=0.72\textwidth]{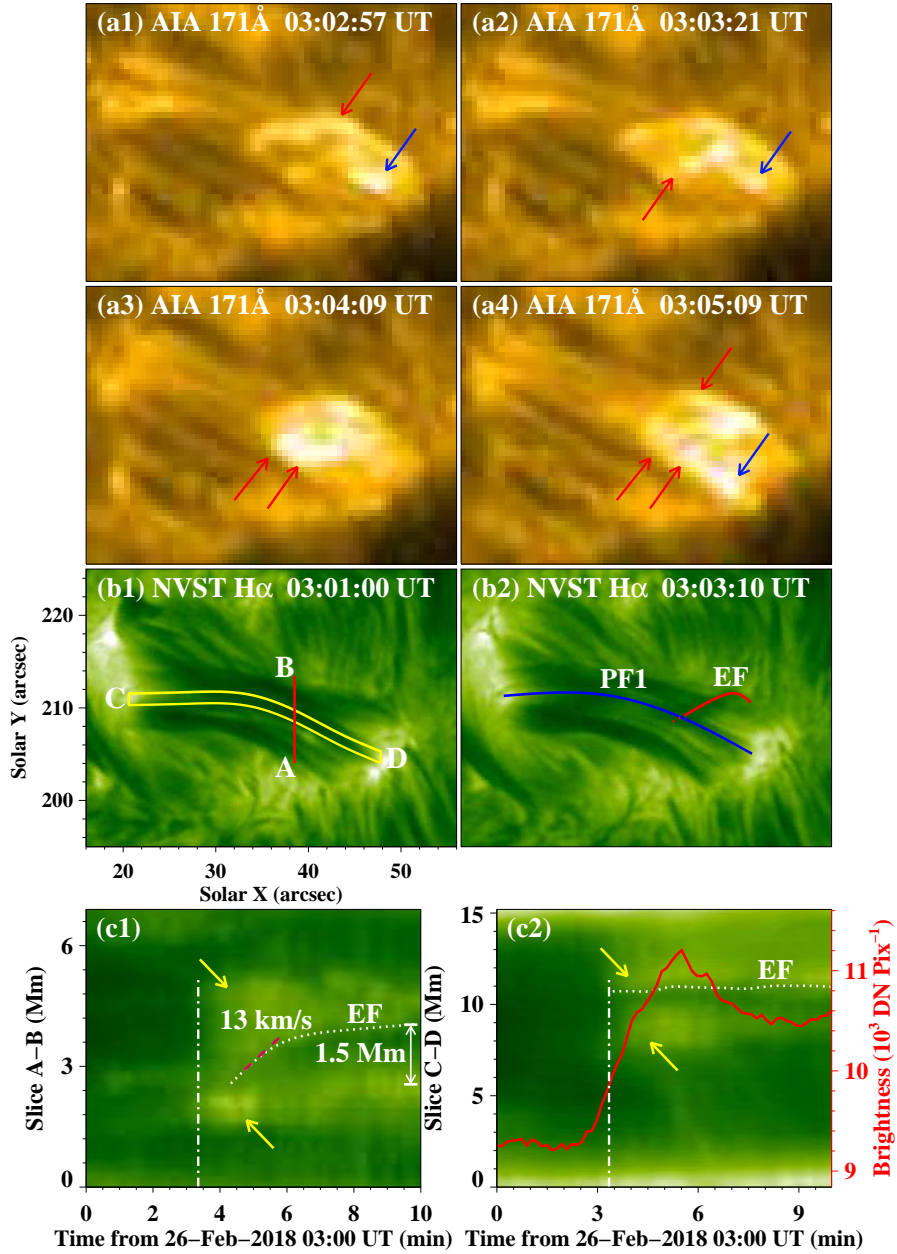}
\caption{Characteristics of the emergence of the small-scale fibrils during the second reconnection scenario.
Panels (a1)-(a4): Brightenings in 171 {\AA} wavelength during the EF emergence. The red arrows denote the EF brightenings that appeared successively. The blue arrows denote the brightening around footpoints of PF1.
Panels (b1)-(b2): NVST H$\alpha$ images acquired before and at the onset of H$\alpha$ brightenings.
The blue and red curves delineate the preexisting fibrils (PF1) and the emerging ones (EF), respectively.
Panels (c1)-(c2): Time-slice plots in H$\alpha$ wavelength along vertical cut ``A-B'' and slit ``C-D" marked in panel (b1), respectively. The white dash-dotted lines mark the onset of the H$\alpha$ brightening. The yellow arrows denote the brightenings in H$\alpha$ wavelength. The red curve is the light curve displaying the variation of the average emission intensity in H$\alpha$ wavelength in slit ``C-D".
}
\label{fig4}
\end{figure}

\begin{figure}
\centering
\includegraphics [width=0.9\textwidth]{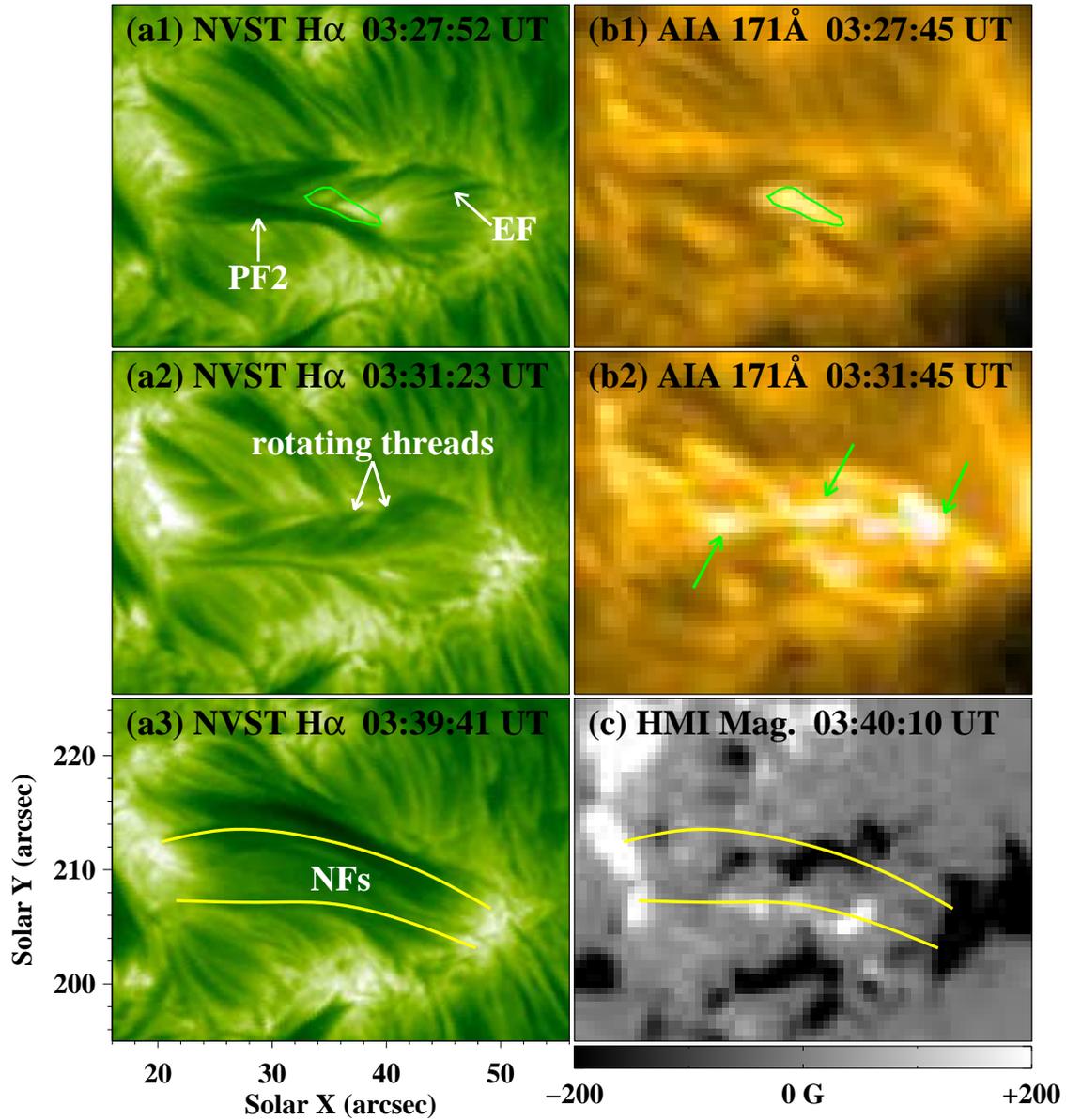}
\caption{Similar to Figures 2-3 but for the third reconnection scenario from 03:18 UT to 03:40 UT.
The green contour and arrows in panels (b1)-(b2) denote the 171 {\AA} brightenings during the reconnection.
Online animation (movie2.mov) displays NVST H$\alpha$ and \emph{SDO}/AIA 171 {\AA} images shown in Figures 3-5. The 17s animation covers ~50 minutes from 02:55 UT to 03:45 UT.
}
\label{fig3}
\end{figure}

\begin{figure}
\centering
\includegraphics [width=0.96\textwidth]{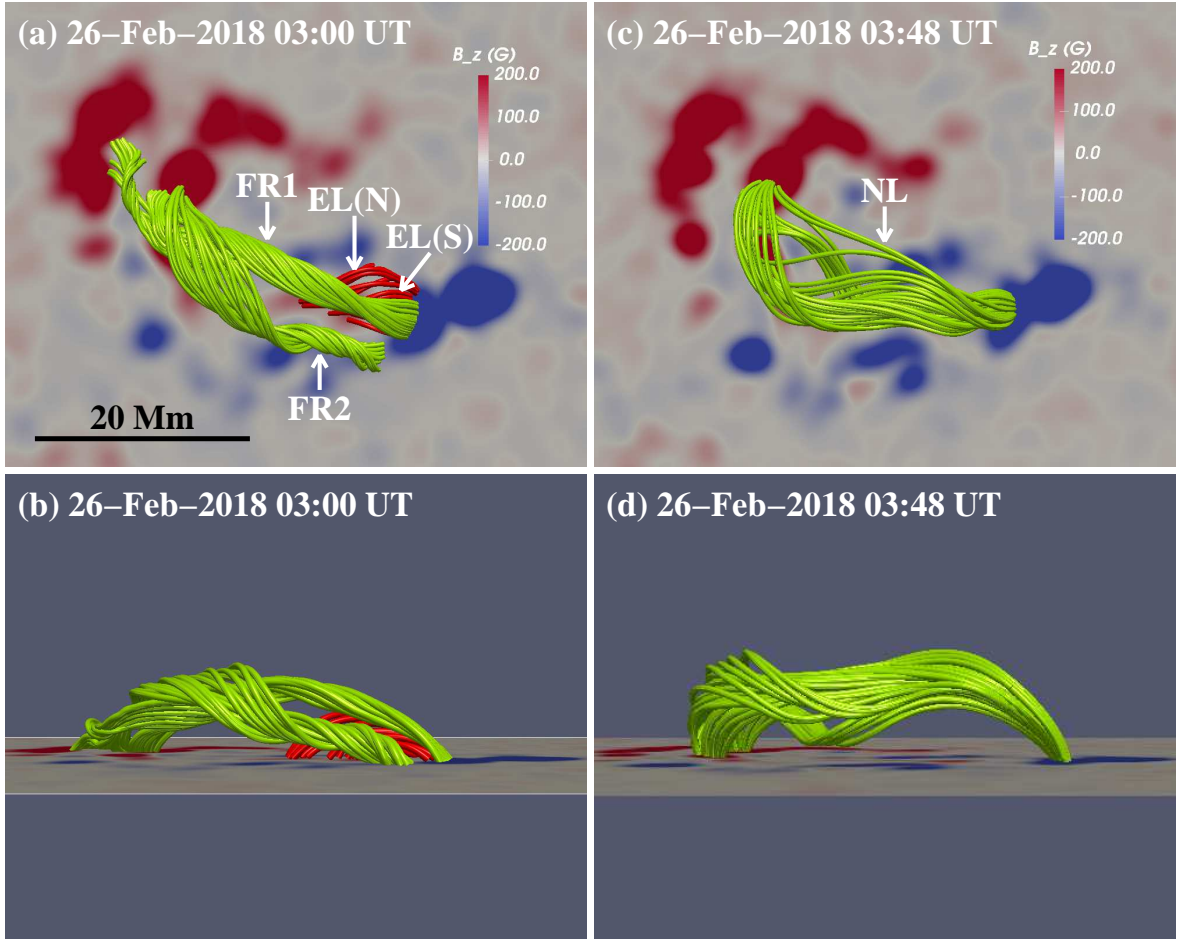}
\caption{Magnetic structures revealed by the NLFFF extrapolations before the second reconnection scenario (panels (a)-(b)) and after the third reconnection scenario (panels (c)-(d)). The photospheric vertical magnetograms (Bz) is shown as the background.
The fist row is the top view of the structure, with north is up and west to the right. The second row is the side view.
The red curves in panels (a)-(b) represent the emerging loops (EL), whose noth part (EL(N)) is higher and south part (EL(S)) is lower.
The two set of green curves represent two magnetic flux ropes (FR1 and FR2).
The green curves in panels (c)-(d) represent the newly-formed twisted loops (NL) resulting from the reconnection.
}
\label{fig5}
\end{figure}


\begin{thebibliography}{}
\bibitem[Archontis \& Hood(2012)]{2012A&A...537A..62A} Archontis, V., \& Hood, A.~W.\ 2012, \aap, 537, A62

\bibitem[Aulanier(2014)]{2014IAUS..300..184A} Aulanier, G.\ 2014, Nature of Prominences and their Role in Space Weather, 300, 184

\bibitem[Baker et al.(2007)]{2007AN....328..773B} Baker, D., van Driel-Gesztelyi, L., \& Attrill, G.~D.~R.\ 2007, Astronomische Nachrichten, 328, 773

\bibitem[Canfield \& Reardon(1998)]{1998SoPh..182..145C} Canfield, R.~C., \& Reardon, K.~P.\ 1998, \solphys, 182, 145

\bibitem[Centeno(2012)]{2012ApJ...759...72C} Centeno, R.\ 2012, \apj, 759, 72

\bibitem[Chen et al.(2017)]{2017ApJ...846..149C} Chen, F., Rempel, M., \& Fan, Y.\ 2017, \apj, 846, 149

\bibitem[Cheung \& Isobe(2014)]{2014LRSP...11....3C} Cheung, M.~C.~M., \& Isobe, H.\ 2014, Living Reviews in Solar Physics, 11, 3

\bibitem[Cheung et al.(2010)]{2010ApJ...720..233C} Cheung, M.~C.~M., Rempel, M., Title, A.~M., \& Sch{\"u}ssler, M.\ 2010, \apj, 720, 233

\bibitem[Chou \& Zirin(1988)]{1988ApJ...333..420C} Chou, D.-Y., \& Zirin, H.\ 1988, \apj, 333, 420

\bibitem[Crooker et al.(2002)]{2002JGRA..107.1028C} Crooker, N.~U., Gosling, J.~T., \& Kahler, S.~W.\ 2002, Journal of Geophysical Research (Space Physics), 107, 1028

\bibitem[De Pontieu et al.(2014)]{2014SoPh..289.2733D} De Pontieu, B., Title, A.~M., Lemen, J.~R., et al.\ 2014, \solphys, 289, 2733

\bibitem[van Driel-Gesztelyi \& Green(2015)]{2015LRSP...12....1V} van Driel-Gesztelyi, L., \& Green, L.~M.\ 2015, Living Reviews in Solar Physics, 12, 1

\bibitem[Edmondson(2012)]{2012SSRv..172..209E} Edmondson, J.~K.\ 2012, \ssr, 172, 209

\bibitem[Edmondson et al.(2010)]{2010ApJ...714..517E} Edmondson, J.~K., Antiochos, S.~K., DeVore, C.~R., Lynch, B.~J., \& Zurbuchen, T.~H.\ 2010, \apj, 714, 517

\bibitem[Fan(2009a)]{2009LRSP....6....4F} Fan, Y.\ 2009a, Living Reviews in Solar Physics, 6, 4

\bibitem[Fan(2009b)]{2009ApJ...697.1529F} Fan, Y.\ 2009b, \apj, 697, 1529

\bibitem[Fan \& Gibson(2004)]{2004ApJ...609.1123F} Fan, Y., \& Gibson, S.~E.\ 2004, \apj, 609, 1123

\bibitem[Galsgaard \& Parnell(2005)]{2005A&A...439..335G} Galsgaard, K., \& Parnell, C.~E.\ 2005, \aap, 439, 335

\bibitem[Gonz{\'a}lez Manrique et al.(2017)]{2017A&A...600A..38G} Gonz{\'a}lez Manrique, S.~J., Bello Gonz{\'a}lez, N., \& Denker, C.\ 2017, \aap, 600, A38

\bibitem[Goode \& Cao(2012)]{2012ASPC..463..357G} Goode, P.~R., \& Cao, W.\ 2012, Second ATST-EAST Meeting: Magnetic Fields from the Photosphere to the Corona., 463, 357

\bibitem[Kong et al.(2018)]{2018ApJ...863L..22K} Kong, D.~F., Pan, G.~M., Yan, X.~L., Wang, J.~C., \& Li, Q.~L.\ 2018, \apjl, 863, L22

\bibitem[Kosugi et al.(2007)]{2007SoPh..243....3K} Kosugi, T., Matsuzaki, K., Sakao, T., et al.\ 2007, \solphys, 243, 3

\bibitem[Kumar et al.(2013)]{2013SoPh..282..503K} Kumar, P., Park, S.-H., Cho, K.-S., \& Bong, S.-C.\ 2013, \solphys, 282, 503

\bibitem[Kumar et al.(2017)]{2017A&A...603A..36K} Kumar, P., Yurchyshyn, V., Cho, K.-S., \& Wang, H.\ 2017, \aap, 603, A36

\bibitem[Lemen et al.(2012)]{2012SoPh..275...17L} Lemen, J.~R., Title, A.~M., Akin, D.~J., et al.\ 2012, \solphys, 275, 17

\bibitem[Li et al.(2014)]{2014A&A...570A..93L} Li, L.~P., Peter, H., Chen, F., \& Zhang, J.\ 2014, \aap, 570, A93

\bibitem[Liu et al.(2014)]{2014RAA....14..705L} Liu, Z., Xu, J., Gu, B.-Z., et al.\ 2014, Research in Astronomy and Astrophysics, 14, 705-718

\bibitem[MacTaggart \& Hood(2010)]{2010ApJ...716L.219M} MacTaggart, D., \& Hood, A.~W.\ 2010, \apjl, 716, L219

\bibitem[Moreno-Insertis(1997)]{1997smf..conf....3M} Moreno-Insertis, F.\ 1997, Solar Magnetic Fields, 3

\bibitem[Moreno-Insertis(2007)]{2007ASPC..369..335M} Moreno-Insertis, F.\ 2007, New Solar Physics with Solar-B Mission, 369, 335

\bibitem[Moore et al.(2002)]{2002mwoc.conf...39M} Moore, R.~L., Falconer, D.~A., \& Sterling, A.~C.\ 2002, Multi-Wavelength Observations of Coronal Structure and Dynamics, 39

\bibitem[Nelson et al.(2013)]{2013MmSAI..84..436N} Nelson, C.~J., Doyle, J.~G., Erd{\'e}lyi, R., Madjarska, M., \& Mumford, S.~J.\ 2013, \memsai, 84, 436

\bibitem[Okamoto et al.(2009)]{2009ApJ...697..913O} Okamoto, T.~J., Tsuneta, S., Lites, B.~W., et al.\ 2009, \apj, 697, 913

\bibitem[Otsuji et al.(2011)]{2011PASJ...63.1047O} Otsuji, K., Kitai, R., Ichimoto, K., \& Shibata, K.\ 2011, \pasj, 63, 1047

\bibitem[Pevtsov et al.(1996)]{1996ApJ...473..533P} Pevtsov, A.~A., Canfield, R.~C., \& Zirin, H.\ 1996, \apj, 473, 533

\bibitem[Pevtsov \& Kazachenko(2004)]{2004ESASP.575..241P} Pevtsov, A.~A., \& Kazachenko, M.\ 2004, SOHO 15 Coronal Heating, 575, 241

\bibitem[Pevtsov et al.(2003)]{2003ApJ...593.1217P} Pevtsov, A.~A., Maleev, V.~M., \& Longcope, D.~W.\ 2003, \apj, 593, 1217

\bibitem[Pesnell et al.(2012)]{2012SoPh..275....3P} Pesnell, W.~D., Thompson, B.~J., \& Chamberlin, P.~C.\ 2012, \solphys, 275, 3

\bibitem[Rempel \& Cheung(2014)]{2014ApJ...785...90R} Rempel, M., \& Cheung, M.~C.~M.\ 2014, \apj, 785, 90

\bibitem[Scherrer et al.(2012)]{2012SoPh..275..207S} Scherrer, P.~H., Schou, J., Bush, R.~I., et al.\ 2012, \solphys, 275, 207

\bibitem[Schmieder et al.(2014)]{2014SSRv..186..227S} Schmieder, B., Archontis, V., \& Pariat, E.\ 2014, \ssr, 186, 227

\bibitem[Schmieder et al.(2013)]{2013A&A...559A...1S} Schmieder, B., Guo, Y., Moreno-Insertis, F., et al.\ 2013, \aap, 559, A1

\bibitem[Shibata et al.(2007)]{2007Sci...318.1591S} Shibata, K., Nakamura, T., Matsumoto, T., et al.\ 2007, Science, 318, 1591

\bibitem[Shibata et al.(1991)]{1991saaj.conf..169S} Shibata, K., Nozawa, S., \& Matsumoto, R.\ 1991, Supercomputing Astronomy and Astrophysics in Japan, 169

\bibitem[Toriumi et al.(2015)]{2015ApJ...811..138T} Toriumi, S., Cheung, M.~C.~M., \& Katsukawa, Y.\ 2015, \apj, 811, 138

\bibitem[Tsuneta et al.(2008)]{2008SoPh..249..167T} Tsuneta, S., Ichimoto, K., Katsukawa, Y., et al.\ 2008, \solphys, 249, 167

\bibitem[Vargas Dom{\'{\i}}nguez et al.(2014)]{2014ApJ...794..140V} Vargas Dom{\'{\i}}nguez, S., Kosovichev, A., \& Yurchyshyn, V.\ 2014, \apj, 794, 140

\bibitem[Vargas Dom{\'{\i}}nguez et al.(2012)]{2012SoPh..278...33V} Vargas Dom{\'{\i}}nguez, S., MacTaggart, D., Green, L., van Driel-Gesztelyi, L., \& Hood, A.~W.\ 2012, \solphys, 278, 33

\bibitem[Wiegelmann(2004)]{2004SoPh..219...87W} Wiegelmann, T.\ 2004, \solphys, 219, 87

\bibitem[Wiegelmann et al.(2006)]{2006SoPh..233..215W} Wiegelmann, T., Inhester, B., \& Sakurai, T.\ 2006, \solphys, 233, 215

\bibitem[Wiegelmann et al.(2012)]{2012SoPh..281...37W} Wiegelmann, T., Thalmann, J.~K., Inhester, B., et al.\ 2012, \solphys, 281, 37

\bibitem[Xue et al.(2016)]{2016NatCo...711837X} Xue, Z., Yan, X., Cheng, X., et al.\ 2016, Nature Communications, 7, 11837

\bibitem[Yan et al.(2017)]{2017ApJ...845...18Y} Yan, X.~L., Jiang, C.~W., Xue, Z.~K., et al.\ 2017, \apj, 845, 18

\bibitem[Zhang \& Low(2003)]{2003ApJ...584..479Z} Zhang, M., \& Low, B.~C.\ 2003, \apj, 584, 479

\bibitem[Zhang \& Low(2001)]{2001ApJ...561..406Z} Zhang, M., \& Low, B.~C.\ 2001, \apj, 561, 406

\bibitem[Zhang et al.(2003)]{2003A&A...399..755Z} Zhang, J., Solanki, S.~K., \& Wang, J.\ 2003, \aap, 399, 755

\bibitem[Zhang \& Wang(2002)]{2002ApJ...566L.117Z} Zhang, J., \& Wang, J.\ 2002, \apjl, 566, L117

\bibitem[Zwaan(1987)]{1987ARA&A..25...83Z} Zwaan, C.\ 1987, \araa, 25, 83


\end{thebibliography}
\end{document}